\documentclass[letter]{aa}
\usepackage{graphicx}
\usepackage{epsfig}
\usepackage{subfigure}
\usepackage{amssymb}
\usepackage{natbib}
\bibpunct{(}{)}{;}{a}{}{,}

\newcommand{\FIG}[1]{#1}
\def\mso{\,{\rm M}_\odot}

 \newcommand{\gcm}{\,{\rm g}\,{\rm cm}^{-3}}
 
 \def\kms{\, {\rm km}\, {\rm s}^{-1}}

 \def\simle{\mathrel{\hbox{\rlap{\hbox{\lower4pt\hbox{$\sim$}}}\hbox{$<$}}}}
 \def\simgr{\mathrel{\hbox{\rlap{\hbox{\lower4pt\hbox{$\sim$}}}\hbox{$>$}}}}

 \def\msoy{\, \mso~{\rm yr}^{-1}}
 \def\pc{\, \mathrm{pc}}

\begin{document}
   \title{A hydrodynamical model of the circumstellar bubble created by two massive stars}

   \author{A. J. van Marle
          \inst{1,2}
          \and
          Z. Meliani
          \inst{3}
          \and
          A. Marcowith
          \inst{4}
          }

   \offprints{A. J. van Marle}

   \institute{Institute of Astronomy, KU Leuven, 
              Celestijnenlaan 200D, B-3001 Heverlee, Belgium \\
              \email{AllardJan.vanMarle@ster.kuleuven.be}
       \and
               Centre for Plasma Astrophysics, Department of Mathematics, KU Leuven, 
              Celestijnenlaan 200B, B-3001 Heverlee, Belgium 
       \and
            Observatoire de Paris, 
            5 place Jules Janssen 92195 Meudon, France \\
              \email{zakaria.meliani@obspm.fr}
       \and
            Laboratoire Univers et Particules (LUPM) Universit{\'e} Montpellier, CNRS/IN2P3, CC72,
            place Eug{\`e}ne Bataillon, F-34095 Montpellier Cedex 5, France \\
            \email{Alexandre.Marcowith@univ-montp2.fr}
}
   \date{Received <date> / Accepted <date>}

\abstract{Numerical models of the wind-blown bubble of massive stars usually only account for the wind of a single star. 
However, since massive stars are usually formed in clusters, it would be more realistic to follow the evolution of a bubble created by several stars.}
{We develope a two-dimensional (2D) model of the circumstellar bubble created by two massive stars, a 40$\mso$ star and a 25$\mso$ star, and follow its evolution. 
The stars are separated by approximately 16 pc and surrounded by a cold medium with a density of 20 particles per cm$^3$.}
{We use the MPI-AMRVAC hydrodynamics code to solve the conservation equations of hydrodynamics on a 2D cylindrical grid using time-dependent models 
for the wind parameters of the two stars. At the end of the stellar evolution (4.5 and 7.0 million years for the 40 and 25$\mso$ stars, respectively), 
we simulate the supernova explosion of each star.}
{Each star initially creates its own bubble. However, as the bubbles expand they merge, creating a combined, aspherical bubble.
The combined bubble evolves over time, influenced by the stellar winds and supernova explosions.}
{The evolution of a wind-blown bubble created by two stars deviates from that of the bubbles around single stars. 
In particular, once one of the stars has exploded, the bubble is too large for the wind of the remaining star to maintain and the outer shell starts to disintegrate. 
The lack of thermal pressure inside the bubble also changes the behavior of circumstellar features close to the remaining star. 
The supernovae are contained inside the bubble, which reflects part of the energy back into the circumstellar medium.}

  \titlerunning{A bubble created by two massive stars}
  \authorrunning{van Marle, Meliani \& Marcowith}


  \maketitle

%

\section{Introduction}
It is well-known that the winds of massive stars create large-scale ($\geq\,10\pc$) bubbles around their progenitors. 
An analytical model of these bubbles was presented by \citet{Weaveretal:1977}. 
Numerical models, which incorporated the time-dependent nature of stellar winds were developed by, for example, 
\citet{GarciaSeguraetal:1996a}, \citet{Freyeretal:2003}, \citet{vanMarleetal:2005}, \citet{Dwarkadas:2007}  and \citet{ToalaArthur:2011}. 
Like the analytical model, these models only considered the evolution of a bubble blown by a single star. 
However, since massive stars are typically formed in star clusters, it is unlikely  that any single star could form a bubble on a scale of tens of parsecs, 
without encountering other massive stars. 
The expanding bubble should instead collide with other, similar bubbles and merge as in the case of \object{WN8}-\object{WN9h} \citep{Mauerhanetal:2010}. 
A good example of the resulting bubble can be found in the \object{Rosetta nebula} around cluster \object{NGC~2244} \citep[e.g.][]{PhelpsLada:1997,LiSmith:2005}.

Because of the numerical problems associated with modeling the bubbles of multiple stars, 
attempts to account for the influence of multiple stars on the evolution of the bubble have been limited to 
the details of the colliding winds of binary stars \citep[e.g.][]{Stevensetal:1992,FoliniWalder:2000,PittardParkin:2010,vanMarleetal:2011b}, 
or dealt with only part of the stellar evolution \citep{Stegeman:2003}. 
\citet{Velazquezetal:2003} simulated the collision between a supernova and an expanding wind bubble, but with a simplified model that assumed the 
supernova progenitor had not previously created a circumstellar bubble through its own wind.

In this paper, we present the result of a two-dimensional (2D) numerical simulation that models the evolution of a bubble, 
blown by the winds of two massive stars and compare these to the bubbles blown by each star individually. 
We follow the evolution of a 25$\mso$ star that passes through the main sequence (MS) and the red supergiant (RSG) phase, and a 40$\mso$ star 
that passes through the MS, RSG, and Wolf-Rayet (WR) phases. 
Since the stars are approximately 16$\pc$ apart, they initially form individual, spherical bubbles 
that expand into the surrounding interstellar medium (ISM).
As the bubbles expand, they collide and merge, forming an aspherical bubble, fed by the winds from both stars. 
Over time, the wind parameters change, altering the morphology of the bubble, until the stars reach the end of their evolution and explode as supernovae. 

We provide movies of the evolution of the bubbles as online material (Appendix~A).

  \begin{table}
 \centering
   \label{tab:stars}
      \caption{
             Wind and supernova parameters, with t$_{\rm end}$ the time from the start of the simulation to the end of each evolutionary phase, 
             $\dot{m}$ the wind mass loss rate, $v_{\rm w}$ the wind veloicty, and
             $M$ and $E$ the total mass and energy injected into the medium. 
              }
      \begin{tabular}{lccccc}
         \hline\hline
         \noalign{\smallskip}
              Phase & t$_{\rm end}$ & $\dot{m}$ & $v_{\rm w}$ & $M$      & $E$  \\
                    &  [\mbox{Myr}]  & [$\msoy$] & [$\kms$]    & [$\mso$] & $10^{47}$[erg] \\
         \noalign{\smallskip}
         \hline
         \noalign{\smallskip}                                                                                                  
          40$\mso$  &              &             &             &          &     \\
         \noalign{\smallskip}
         \hline
         \noalign{\smallskip}                                                                                                  
          MS       & 4.3 & 1.0$\times10^{-6}$ & 2\,000  & 4.3  & 1\,710    \\
          RSG      & 4.5 & 5.0$\times10^{-5}$ & 15      & 10.0 & 0.224     \\
          WR       & 4.77 & 1.0$\times10^{-5}$ & 2\,000  & 3.0  & 1\,190    \\
          SN       &     &                    &         & 10.0 & 10\,000   \\
         \noalign{\smallskip}
         \hline
         \noalign{\smallskip}                                                                                                  
          25$\mso$  &              &             &             &          &     \\
         \noalign{\smallskip}
         \hline
        \noalign{\smallskip}  
           MS        & 6.5 & 2.0$\times10^{-7}$ & 1\,500 & 1.3  & 170      \\
           RSG       & 7   & 1.0$\times10^{-5}$ & 15     & 5.0  & 0.112    \\
           SN        &     &                    &        & 7.0  & 10\,000  \\
          \noalign{\smallskip}
         \hline
      \end{tabular}
   \end{table}

\section{Numerical setup}
We use the {\tt MPI-AMRVAC} hydrodynamics code \citep{Keppensetal:2012}, 
which solves the conservation equations for mass, momentum, and energy on an adaptive mesh grid. 
The effect of radiative cooling is included \citep{vanMarleKeppens:2011}, 
using a cooling curve for solar-metallicity gas generated with the {\tt CLOUDY} code (Ferland et al. 1998; Wang Ye, private communication).
We set a minimum temperature of 100\,K throughout the simulation, which limits the amount of compression due to radiative cooling, preventing numerical problems. 
Here we neglect the effect of photoionization, 
which would increase the gas temperature within the Str{\"o}mgren radius to about 10,000\,K. 

We set up our simulation as a 2D cylindrically symmetric grid in the $r,z$-plane. 
The grid has a basic resolution of 120$\times$240 gridpoints over 100$\times$200$\pc$. 
The adaptive mesh has four layers, allowing for an effective resolution of 960$\times$1920 gridpoints. 
At the start of the simulation, this grid is filled with gas with a constant density of $10^{-22.5}\gcm$ and solar metallicity.
The stars are defined as small spheres ($R=1.5\pc$), placed 16.2$\pc$ apart that are filled, at the beginning of each timestep, 
with material according to the wind parameters given in Table~\ref{tab:stars}. 
The duration of the evolutionary phases and the mass-loss rates are based on the parameters for a 25$\mso$ star and a 40$\mso$ star 
used by \citet{vanMarleetal:2004} and \citet{vanMarleetal:2005}, respectively.
Wind velocities reflect those typically found in observations \citep[][and references therein]{Lamerscassinelli:1999}. 
For the 40$\mso$ star, the mass-loss rate in the final phase (the Wolf-Rayet phase) has been reduced to reflect the values obtained by \citet{Vinkdekoter:2005}. 

When a star reaches the end of its evolution, 
we fill a sphere ($R=0.75\pc$) with a constant density and thermal energy according to the supernova values given in Table~\ref{tab:stars}. 
The masses reflect the final masses of the stars as well the as values calculated by, e.g., \citet{EldridgeTout:2004}. 
For simplicity, we use a supernova energy of $10^{51}$erg, which is typical of core-collapse supernovae \citep[e.g.][and references therein]{vanMarleetal:2010}

\begin{figure}
\FIG{
 \centering
\mbox{
\includegraphics[width=0.49\textwidth]{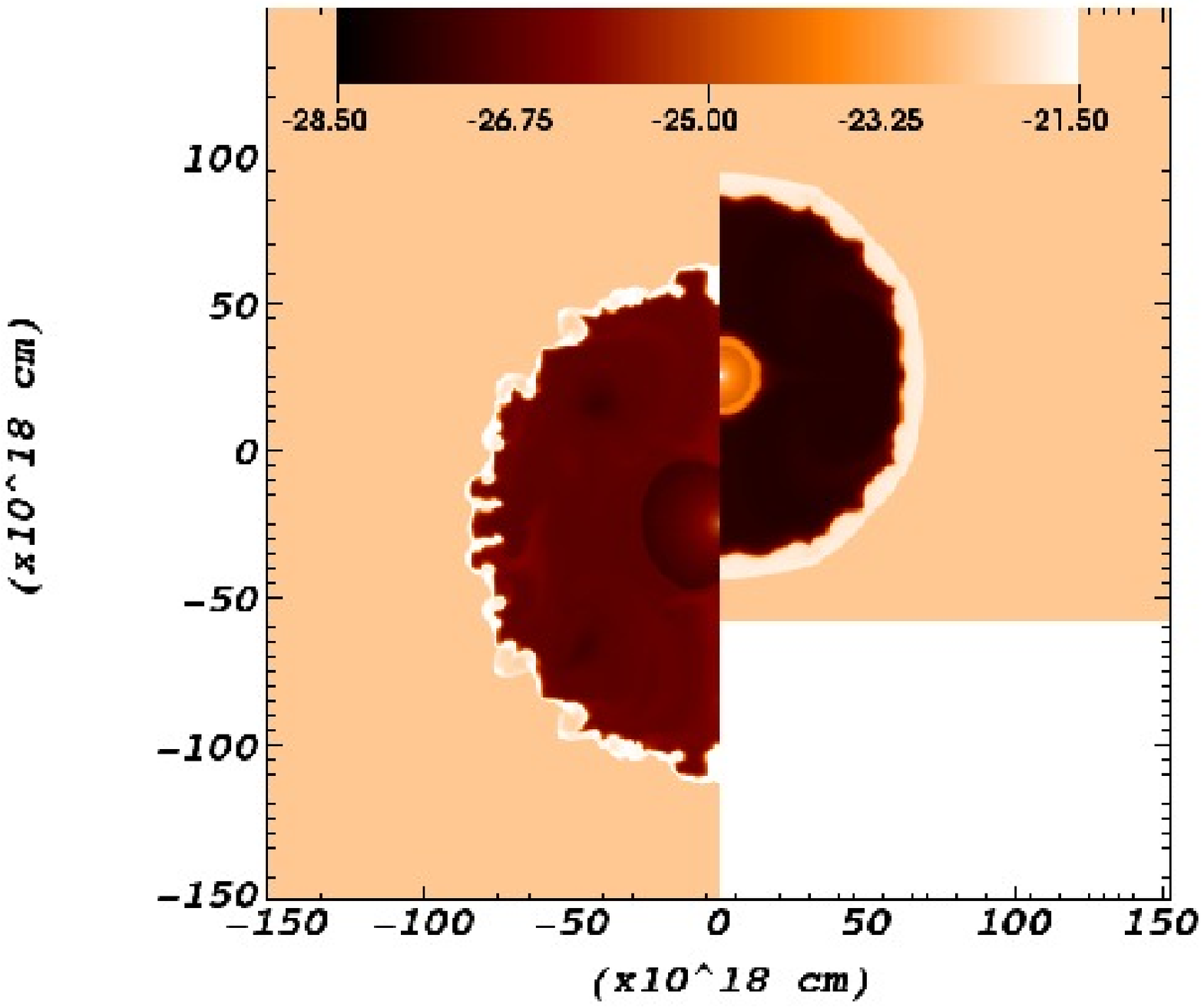}}
}
\caption{Logarithm of the density of the circumstellar medium in g/cm$^3$ at the end of the evolution of the 40$\mso$ star (left) and the 25$\mso$ star (right), 
if the two bubbles never merge. 
}
 \label{fig:star1_2}
\end{figure}

\begin{figure*}
\FIG{
 \centering
\mbox{
\subfigure
{\includegraphics[width=0.32\textwidth]{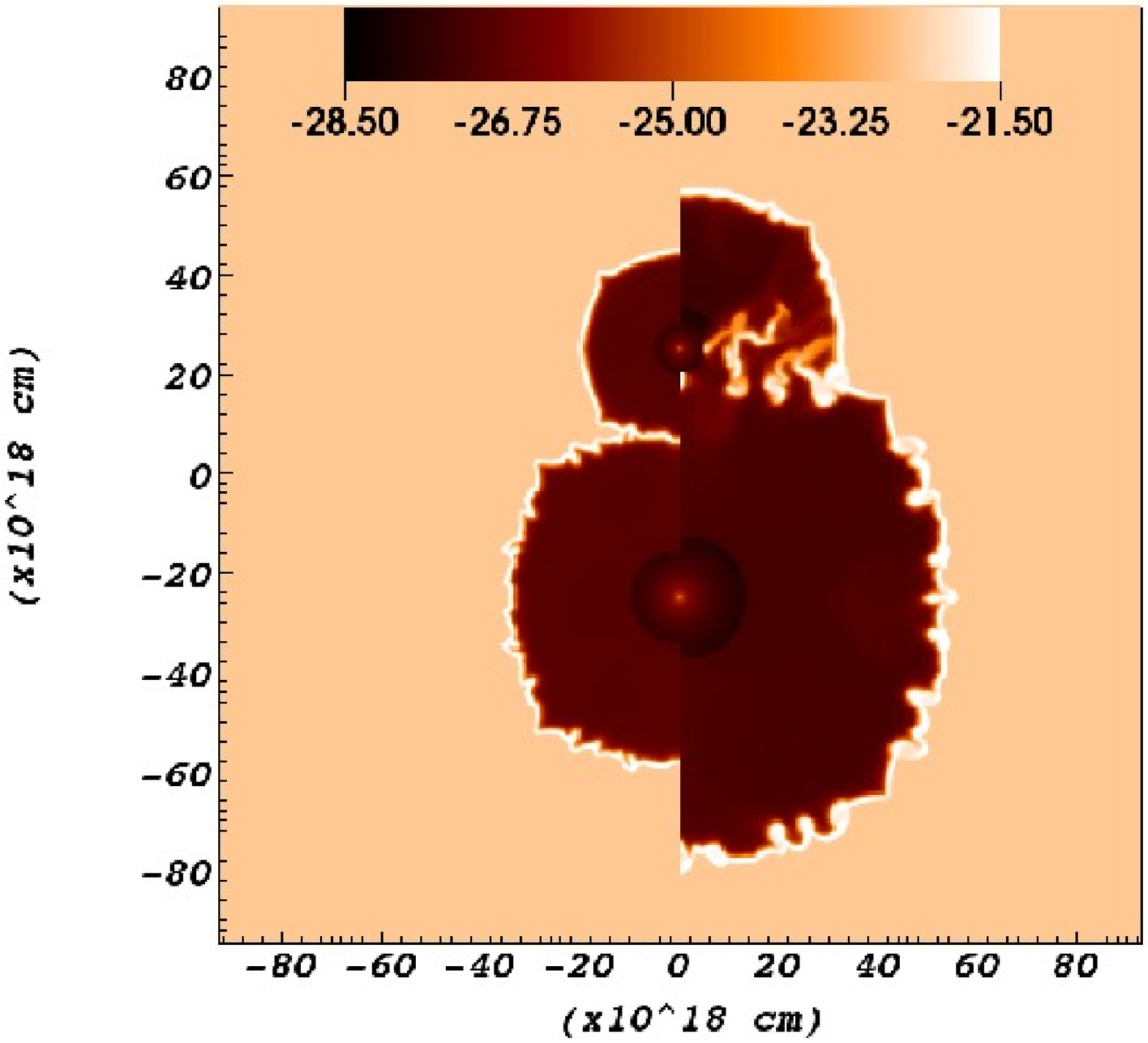}}
\subfigure
{\includegraphics[width=0.32\textwidth]{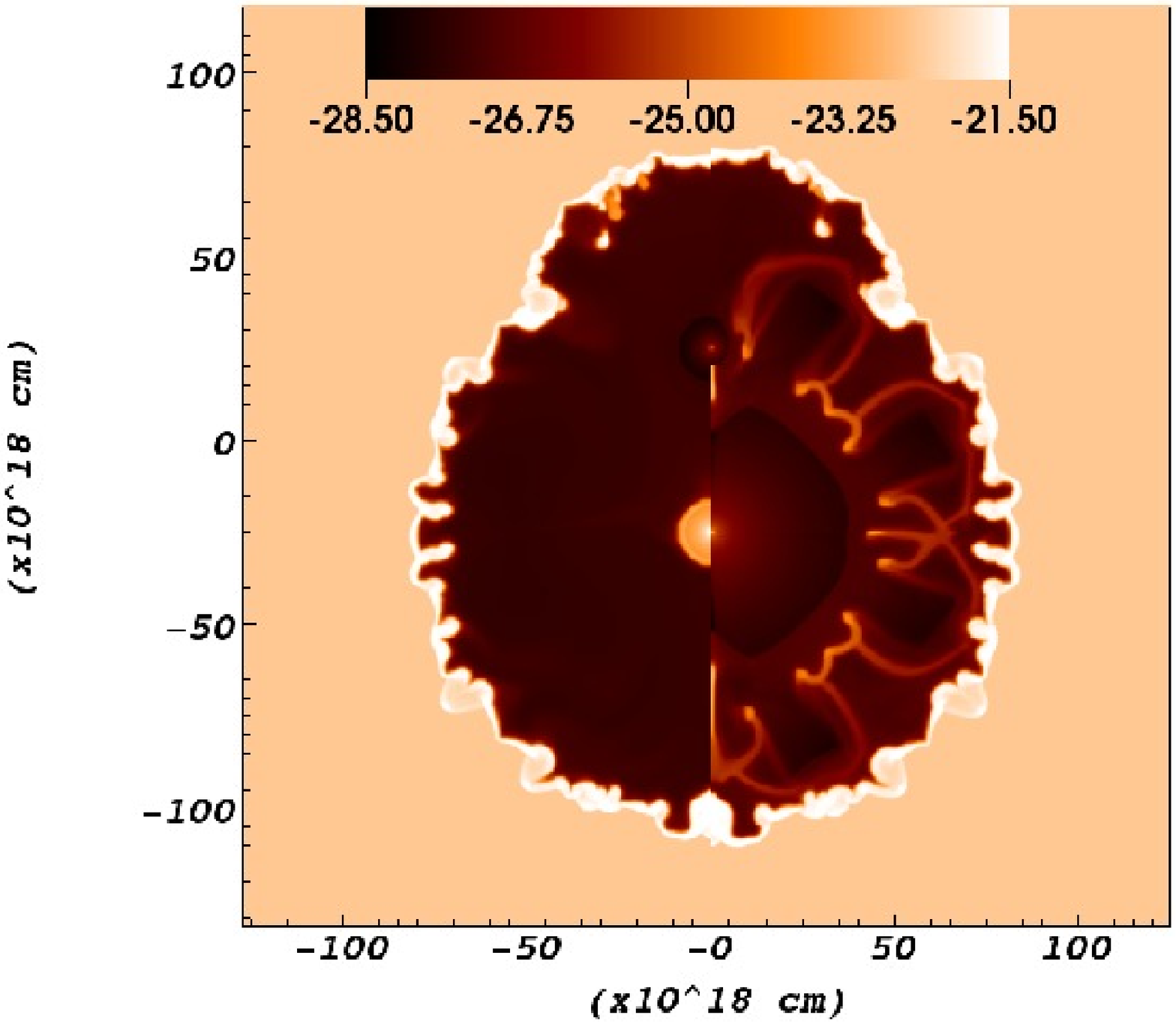}}
\subfigure
{\includegraphics[width=0.32\textwidth]{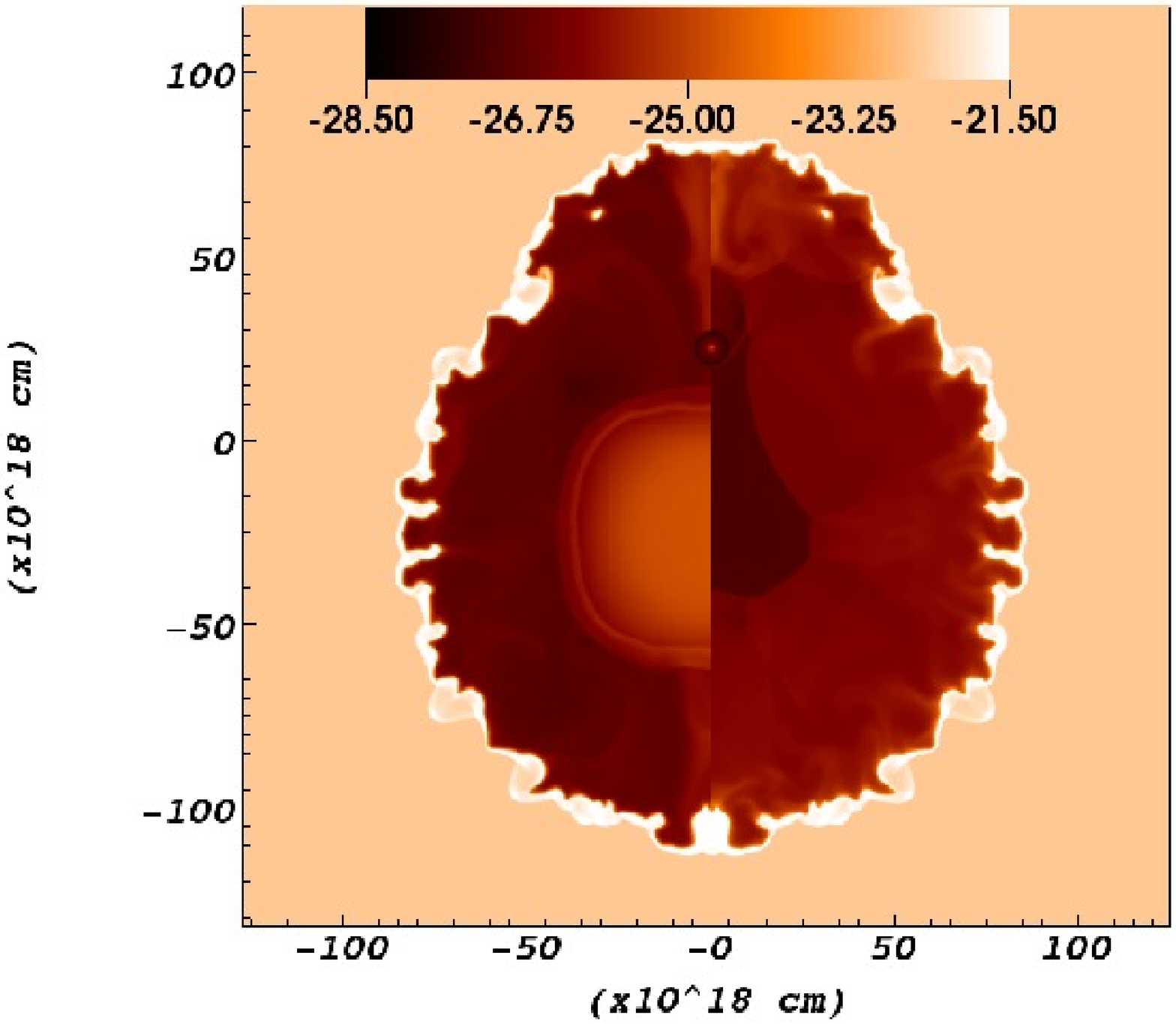}}
}
}
\caption{Logarithm of the density of the circumstellar medium in g/cm$^3$ after 1.17 and 2.35 Myr (left panel), 
after 4.42 and 4.58 Myr (center panel), and after 4.774 and 4.783 Myr (right panel). 
On the left, the shells of the two bubbles encounter each other (left panel, left side) and are destroyed in the process. 
The remnants are pushed toward the lightest of the stars (left panel, right side). 
On the right, the 40$\mso$ star (at $Y=-2.5\times20^{19}$cm) goes through the RSG phase, 
which forms a thin shell at the wind termination  shock (center panel, left side) and then the WR phase. 
The WR wind sweeps up the RSG wind and the RSG shell is destroyed in the collision. 
The debris of the collision moves outward into the hot shocked gas (central panel, right side). 
Finally, the 40$\mso$ star explodes as a supernova, which expands into the circumstellar
bubble (right panel, left). 
Once it has hit the outer shell, the supernova bounces back and starts 
flowing inward (right panel, right).
}
 \label{fig:firsthalf}
\end{figure*}

\begin{figure*}
\FIG{
 \centering
\mbox{
\subfigure
{\includegraphics[width=0.32\textwidth]{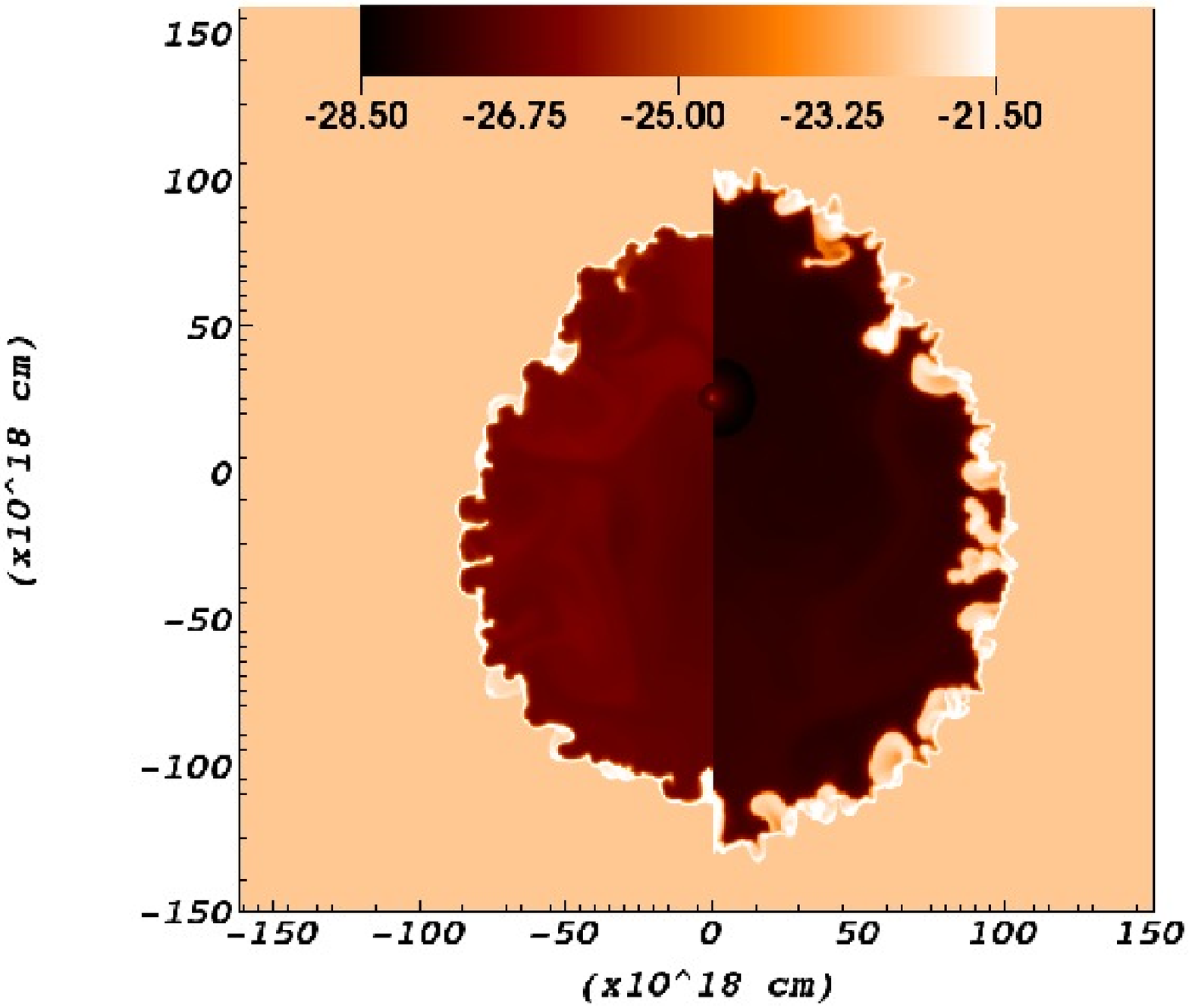}}
\subfigure
{\includegraphics[width=0.32\textwidth]{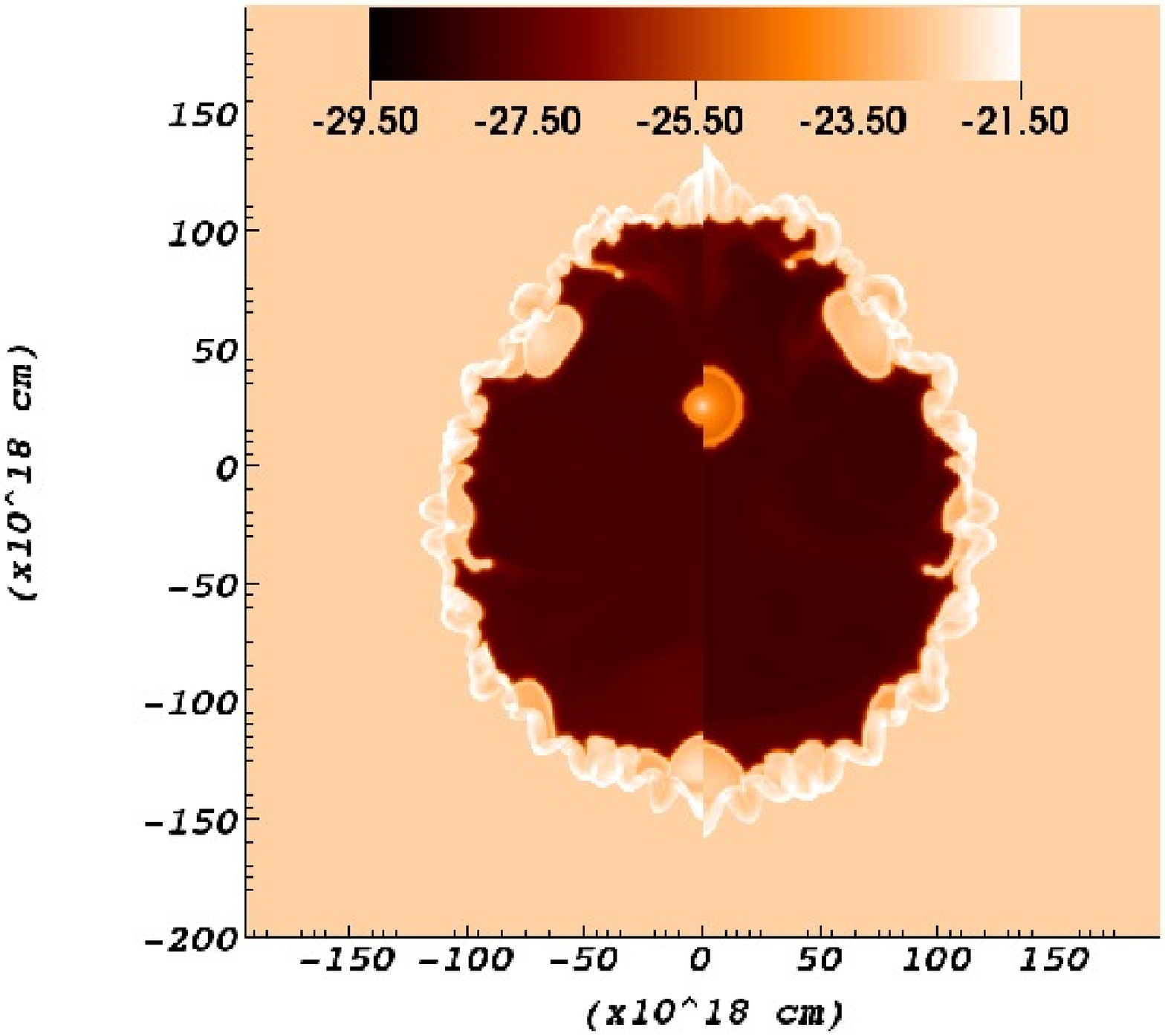}}
\subfigure
{\includegraphics[width=0.32\textwidth]{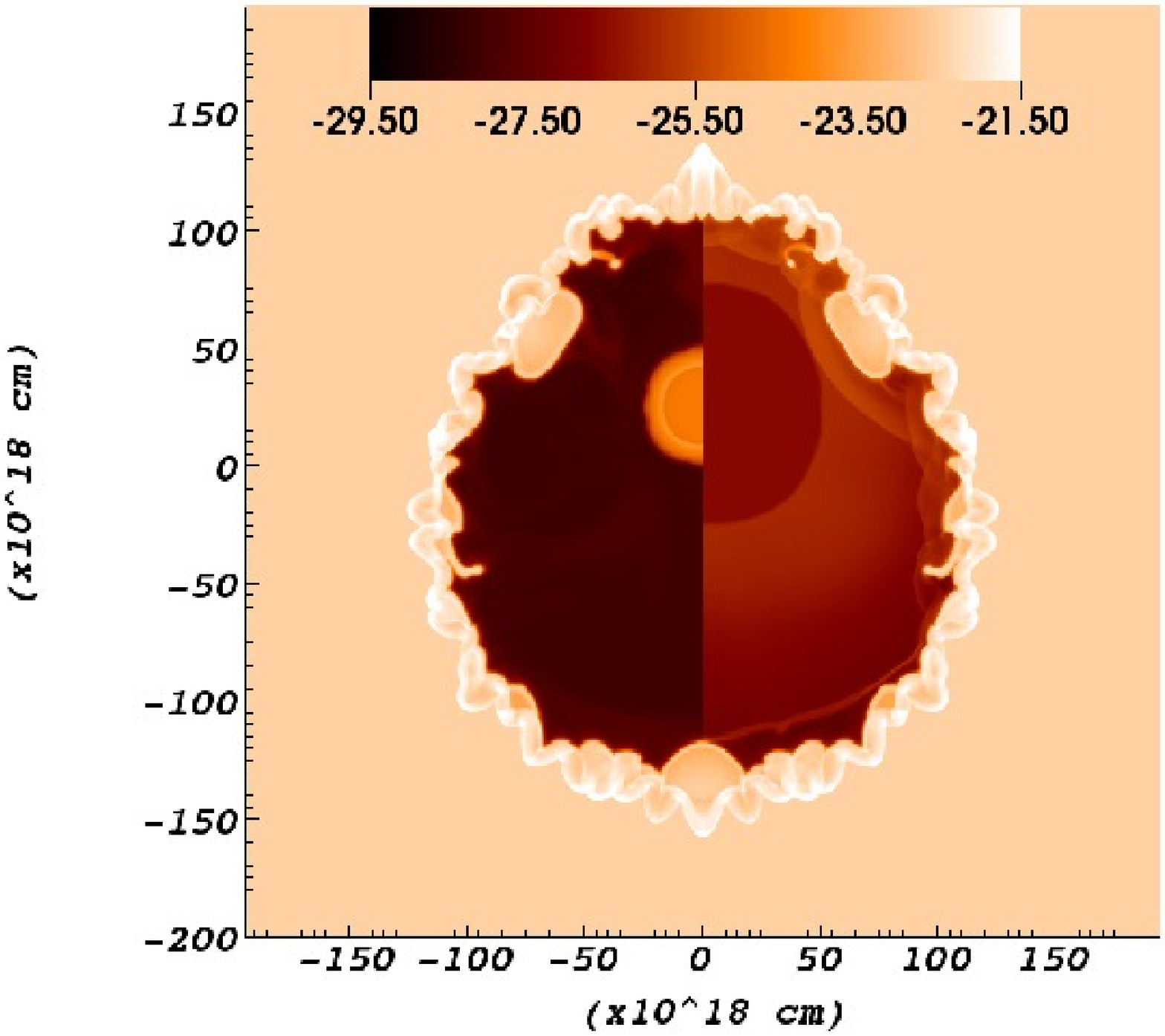}}
}
}
\caption{Logarithm of the density of the circumstellar medium in g/cm$^3$ after 4.86 and 5.60 Myr (left panel), 
after 6.58 and 7.0 Myr (center panel), and after 7.01 and 7.07 Myr (right panel). 
On the left, the mass, ejected by the first supernova has spread throughout the bubble, which, due to the injection of energy, expands rapidly. 
The density inside the bubble decreases and the Rayleigh-Taylor instabilities in the outer shell become more pronounced. 
In the central panel, the 25$\mso$ star has reached the RSG phase and forms a shell at the wind termination shock that moves away from the star as the 
thermal pressure in the bubble decreases over time. 
Finally, on the right, the 25$\mso$ star explodes as a supernova, which sweeps up the RSG shell and fills the bubble. 
}
 \label{fig:postsn1}
\end{figure*}

\section{Results}
\subsection{Single stars}
For comparison, we run the evolution of the circumstellar bubble for each star individually. 
Fig.~\ref{fig:star1_2} shows the density of the circumstellar medium at the end of the evolution of each star. 
The 40$\mso$ star has created a large bubble (R$\sim25$\,pc) that has swept up a thin, 
shell containing linear thin-shell instabilities \citep{Vishniac:1983}. 
The star has already passed through the RSG phase to become a WR star, 
and the WR wind has swept up the remnants of the earlier phases \citep{GarciaSeguraetal:1996b,vanMarleetal:2005}. 
The 25$\mso$ star ends its evolution as a RSG and forms a thin shell at the wind termination shock (~$\sim3$\,pc). 
The total bubble is smaller (R$\sim23$\,pc) than for the 40$\mso$ star, 
though the difference is small because the longer life of the smaller star partially compensates for the lower ram pressure of its wind.  
The swept-up shell at the outer edge is much thicker. 
This is the result of the lower ram pressure of the wind, which causes the bubble to expand at a lower rate, 
giving the swept-up shell more time to expand under its own internal pressure. 
Because the shell is thicker, it is not subject to thin-shell instabilities and only shows the beginning of Rayleigh-Taylor instabilities at its inner edge.

\subsection{Two stars combined}
Each star initially creates its own circumstellar bubble.  
These bubbles follow the pattern predicted by \citet{Weaveretal:1977}: The wind expands freely until it hits the termination shock. 
The kinetic energy of the wind is then turned into thermal energy creating a volume filled with very hot ($T\sim10^7-10^8$K) shocked wind material. 
The high thermal pressure of the shocked wind material causes it to expand, sweeping up a shell of shocked ISM, which moves outward. 
Eventually, the swept-up shells of the two bubbles encounter each other and both are fractured during the collision (left panel of Fig.~\ref{fig:firsthalf}). 
The remnants are pushed into the bubble with the lowest thermal pressure (resulting from the weakest wind, and therefore, typically, the lightest star). 
Once the shells have been fractured, the two bubbles merge, creating a single, 
aspherical bubble in which the thermal pressure is equalized as the shocked gas of the two stellar winds mixes. 
The thin outer shell is unstable and shows both linear thin-shell \citep{Vishniac:1983} and Rayleigh-Taylor instabilities. 

The 40$\mso$ star, which has the shortest lifespan, leaves the main sequence and goes through the RSG and WR phases (central panel of Fig.~\ref{fig:firsthalf}).  
The RSG wind forms a thin shell at the termination shock, which is destroyed by the WR wind.
Finally, the star explodes as a supernova. 
At this time, the bubble is still quite aspherical, despite the outer shell being driven by the thermal pressure in the shocked wind. 
It has a maximum length (along the axis connecting the stars) of 57\,pc. 
At the position of the 40$\mso$ star, it is $\sim$52\,pc wide, whereas at the position of the $25\mso$ star the width is only 39\,pc. 
The shape of the outer shell still indicates that it is the merger of two spherical bubbles. 
 
The supernova quickly expands into the low density bubble that has been created by the stellar winds, 
but stops when it reaches the contact discontinuity separating the shocked wind from the shocked ISM (right panel of Fig.~\ref{fig:firsthalf}).
The injection of energy from the supernova (approximately three times the total energy injected so far by the stellar winds) 
causes the bubble to expand more  rapidly (left panel of Fig.~\ref{fig:postsn1}). 
Because the supernova has injected a lot of energy, but (relatively) little mass, the density inside the bubble decreases quickly during the expansion 
and the Rayleigh-Taylor instabilities in the outer shell grow quickly. 
The bubble is left with only one source of energy, the wind from the 25$\mso$ star, which is insufficient to maintain the 
expansion of the bubble. 
As a result, the thermal pressure inside the wind bubble decreases leading to a loss of compression at the outer shell, 
which starts to expand because of its own internal pressure (left panel of Fig.~\ref{fig:postsn1}).  
During this phase, the asymmetrical shape of the bubble starts to disappear as the isotropic pressure in the shocked wind bubble pushes 
the shell outward equally in all directions. 

Eventually, the 25$\mso$ star reaches the RSG phase, which further reduces the energy being added to the bubble. 
A thin shell forms at the wind termination shock (central panel of Fig.~\ref{fig:postsn1}). 
This initially happens close to the star because of the low ram pressure of the slow RSG wind. 
However, owing to the relatively low pressure in the bubble, this shell can move farther away from the star than would have been possible if the 25$\mso$ star 
had been alone ($R\sim 6.5\,$pc, rather than $R\sim 4.8\,$pc when comparing the right side of the center panel of Fig.~\ref{fig:postsn1} with Fig~\ref{fig:star1_2}).

Finally, the 25$\mso$ star reaches the end of its evolution and explodes (right panel of Fig.~\ref{fig:postsn1}), 
sweeping up the surrounding RSG wind as well as the RSG shell. 
The bubble has lost most of its asymmetry; it has a length of $\sim$105\,pc and a maximum width of $\sim$104\,pc.
The outer edge of the shell has an almost spherical shape. 
As in the case of the first star, 
the supernova quickly expands into the low density shocked wind bubble until it reaches the outer shell, then falls back inward. 
From this moment on, the bubble has no energy source. 
While energy continues to be lost from the bubble by thermal conduction, turbulent mixing, and radiative cooling, 
the outer shell starts to push inward owing to is own internal pressure, which is considerable because of its relatively high density. 
Assuming a typical speed for gas in the shell of 1...2\,$\kms$ (the sound speed for gas at a $10^2-10^3$\,K), 
the gas in the outer shell would take a minimum of $2.5\times10^7$\,years to reach the center,  
though in practice it would take longer owing to the continuing resistance of the hot bubble.

\section{Discussion and conclusions}
The presence of multiple stars inside the circumstellar bubble influences the evolution of the bubble in several ways. 

The most obvious is in terms of the outer shape of the bubble, which is strongly aspherical during the early phase of its evolution. 
This effect is clearly visible until after the supernova explosion of the first star.  
In addition, remnants of the collision between the two outer shells of the individual bubbles remain visible within the shocked wind bubble 
until after the supernova explosion of the 40$\mso$ star.

The collapse of the outer shell, which is caused by a decrease in the internal pressure of the bubble, is unique to the multi-star scenario. 
It can only occur when the bubble persists for an extended period of time ($\sim10^5-10^6$ yr) with a significantly diminished energy source 
(the wind of the remaining star). 
The effective result is that the outer shell resembles the single-star shell of the star with a stronger wind. 
As long as the 40$\mso $ star exists, the shell looks like a shell created by a single 40$\mso$ star. 
Afterwards, it starts to resemble the circumstellar shell of a 25$\mso$ star as it becomes relatively thick. 
However, it is far more instable as a result of the instabilities created in the initial evolution.
This lack of internal pressure also affects the medium close to the  25$\mso$ star: 
the RSG shell can move further away from its progenitor than would have been the case for a single-star bubble. 

The supernovae are contained by the outer shell and do not break out of the bubble. 
This is surprising, considering the amount of energy in the supernova. 
However, the mass of the outer shell is very high. 
A bubble with $R=30$\,pc and an ambient medium density of $10^{-22.5}\gcm$ has a 5.3$\times10^4\,\mso$ outer shell 
that forms a very effective barrier against a supernova, which is high in energy, but low in mass.

We neglected the effect of photoionization, which would give the gas a minimum temperature of about 10,000\,K during the main sequence and WR phases. 
During these phases, the shells would become thicker owing to their higher thermal pressure. 
This increase in pressure within the outer shell would also hasten its expansion after the first supernova.

We conclude that simulating the bubbles around more than one star improves our understanding of the evolution of the circumstellar bubbles of massive stars.

\begin{acknowledgements} 
A.J.v.M.\ acknowledges support from FWO, grant G.0277.08, K.U.Leuven GOA/2008/04 and GOA/2009/09. 
We thank Dr. Wang Ye at the Department of Physics \& Astronomy, University of Kentucky, for providing us with the radiative cooling curve. 
\end{acknowledgements}

\bibliographystyle{aa}
\bibliography{vanmarle_biblio}

\IfFileExists{vanmarle_biblio.bbl}{}
 {\typeout{}
  \typeout{******************************************}
  \typeout{** Please run "bibtex \jobname" to obtain}
  \typeout{** the bibliography and then re-run LaTeX}
  \typeout{** twice to fix the references!}
  \typeout{******************************************}
  \typeout{}
 }

\listofobjects

\Online
\begin{appendix}  
\section{Animations of the evolution of circumstellar bubbles}

\begin{figure*}
\caption{
The first movie (star1) shows the evolution of the bubble created by the 40$\mso$ star on its own. 
It follows the evolution from the main sequence, to the RSG stage, and then onto the end of the WR phase.
The interval between frames is 39\,112\, years.}
\end{figure*}

\begin{figure*}
\caption{
The second movie (star2) follows the evolution of the bubble created by the 25$\mso$ star on its own. 
It follows the evolution to the end of the RSG phase.
The interval between frames is 391\, years.} 
\end{figure*}

\begin{figure*}
\caption{
The third movie (wwsn$\_$presn1) shows the evolution of the bubble prior to the first supernova. 
The 40$\mso$ star evolves through the main sequence, the red supergiant phase, and then the Wolf-Rayet phase. 
The 25$\mso$ star remains on the main sequence. 
As the two bubble merge, the shells collide and break up. The remnants are pushed toward the star with the lowest mass, because it has the weakest wind.
The interval between frames is 39\,112\, years.}
\end{figure*}

\begin{figure*}
\caption{
The fourth movie (wwsn$\_$sn1) shows the expansion of the first supernova. 
The 40$\mso$ star explodes as a supernova, which expands within the wind-blown bubble of the two stars. 
After colliding with the outer shell, the supernova remnant falls back toward the center of the bubble. 
The interval between frames is 391\, years.} 
\end{figure*}

\begin{figure*}
\caption{
The fifth movie (wwsn$\_$postn1) shows the evolution of the bubble between the two supernovae. 
The second star evolves through the remaining main sequence phase and the red supergiant phase. 
The bubble, which  no longer derives energy from the two stars, loses the thermal pressure necessary to sweep up the outer shell, which starts to fall apart. 
The interval between frames is 39\,112\, years.}
\end{figure*}

\begin{figure*}
\caption{
The sixth movie (wwsn$\_$sn2) shows the expansion of the second supernova. 
The second star explodes as a supernova, which expands inside the bubble.
The interval between frames is 391\, years.}
\end{figure*}

\end{appendix}
\end{document}